\documentclass[letterpaper, 10 pt, conference]{ieeeconf}  % Comment this line out if you need a4paper
\usepackage{amsmath}
\usepackage{amsfonts}
\usepackage{mathtools}
\usepackage{bbold}
\usepackage{graphicx}
\usepackage{color}
\usepackage{tikz}
\usetikzlibrary{positioning}
\usepackage{tikzscale}
\usepackage{import}
\usetikzlibrary{backgrounds,calc,shadings,shapes.arrows,arrows,shapes.symbols,shadows,positioning,decorations.markings,backgrounds,arrows.meta}
\usepackage{caption}
\usepackage{stackengine}
\usepackage{hyperref}

\usepackage{array}
\usepackage{booktabs}
\usepackage{multirow}
\usepackage{siunitx}
\usepackage{arydshln}
\makeatletter
\def\adl@drawiv#1#2#3{%
        \hskip.5\tabcolsep
        \xleaders#3{#2.5\@tempdimb #1{1}#2.5\@tempdimb}%
                #2\z@ plus1fil minus1fil\relax
        \hskip.5\tabcolsep}
\newcommand{\cdashlinelr}[1]{%
  \noalign{\vskip\aboverulesep
           \global\let\@dashdrawstore\adl@draw
           \global\let\adl@draw\adl@drawiv}
  \cdashline{#1}
  \noalign{\global\let\adl@draw\@dashdrawstore
           \vskip\belowrulesep}}
\newcommand{\cdashlinelrdotted}[1]{%
  \noalign{\vskip\aboverulesep
           \global\let\@dashdrawstore\adl@draw
           \global\let\adl@draw\adl@drawiv}
  \cdashline{#1}[.4pt/1pt]
  \noalign{\global\let\adl@draw\@dashdrawstore
           \vskip\belowrulesep}}
\makeatother

\definecolor{darkblue}{HTML}{1f4e79}
\definecolor{lightblue}{HTML}{00b0f0}
\definecolor{salmon}{HTML}{ff9c6b}
\definecolor{dodgerblue}{rgb}{0.12, 0.56, 1.0}
\definecolor{frenchblue}{rgb}{0.0, 0.45, 0.73}
\definecolor{green(pigment)}{rgb}{0.0, 0.65, 0.31}
\definecolor{macaroniandcheese}{rgb}{1.0, 0.74, 0.53}
\definecolor{arylideyellow}{rgb}{0.91, 0.84, 0.42}
\definecolor{pansypurple}{rgb}{0.47, 0.09, 0.29}
\definecolor{glaucous}{rgb}{0.38, 0.51, 0.71}
\definecolor{hanblue}{rgb}{0.27, 0.42, 0.81}
\definecolor{newblue}{rgb}{0.56, 0.67, 0.85}
\definecolor{newgreen}{rgb}{0.67, 0.82, 0.57}
\definecolor{fireenginered}{rgb}{0.81, 0.09, 0.13}

\IEEEoverridecommandlockouts                              % This command is only needed if 
\title{\LARGE \bf
Stochastic Graph Heat Modelling\\ for Diffusion-based Connectivity Retrieval
}

\author{Stephan Goerttler\authorrefmark{1}\authorrefmark{2}$^{1}$, Fei He\authorrefmark{1} and Min Wu\authorrefmark{2}\\
\authorrefmark{1}Centre for Computational Science and Mathematical Modelling, Coventry University, Coventry, UK\\
\authorrefmark{2}Institute for Infocomm Research, A*STAR, Singapore% <-this % stops a space
\thanks{$^{1}$Stephan Goerttler (goerttlers@uni.coventry.ac.uk) is supported by the A*STAR Research Attachment Program (ARAP).}}
\begin{document}
%\bstctlcite{IEEEexample:BSTcontrol}
%\mathtoolsset{showonlyrefs,showmanualtags}

\maketitle
\thispagestyle{empty}
\pagestyle{empty}

%%%%%%%%%%%%%%%%%%%%%%%%%%%%%%%%%%%%%%%%%%%%%%%%%%%%%%%%%%%%%%%%%%%%
\begin{abstract}
Heat diffusion describes the process by which heat flows from areas with higher temperatures to ones with lower temperatures. This concept was previously adapted to graph structures, whereby heat flows between nodes of a graph depending on the graph topology. 
%Here, we extend the graph heat equation to include stochasticity, 
Here, we combine the graph heat equation with the stochastic heat equation,
which ultimately yields a model for multivariate time signals on a graph. We show theoretically how the model can be used to directly compute the diffusion-based connectivity structure from multivariate signals. 
%We compare our method to other measures of connectivity, namely pairwise Pearson correlation and smooth signal correlation, which incidentally build on ideas related to diffusion. 
Unlike other connectivity measures, our heat model-based approach is inherently multivariate and yields an absolute scaling factor, namely the graph thermal diffusivity, which captures the extent of heat-like graph propagation in the data.
On two datasets, we show how the graph thermal diffusivity can be used to characterise Alzheimer's disease (AD). We find that the graph thermal diffusivity is lower for AD patients than healthy controls and correlates with mini–mental state examination (MMSE) scores, suggesting structural impairment in patients in line with previous findings.
\newline
\indent \textit{Clinical relevance}— This study introduces a novel heat-based connectivity measure, which allows to characterise Alzheimer's disease in terms of the graph thermal diffusivity.
\end{abstract}

%%%%%%%%%%%%%%%%%%%%%%%%%%%%%%%%%%%%%%%%%%%%%%%%%%%%%%%%%%%%%%%%%%%%%%%%%%%%%%%%
\section{Introduction}
%xxx include graphwave \cite{donnat2018learning}
%xxx include atasoy connectome \cite{atasoy2016human} simulation
%\cite{goerttler2022effect}
Diffusion encompasses the general notion that connected areas with differing concentrations, for example chemical concentrations or heat densities, are aimed at equalling each other out over time \cite{crank1979mathematics}. While diffusion processes are complex on a microscopic level, their macroscopic description often reduces to partial differential equations, such as the heat equation. 
%To model chaotic microscopic effects
This equation can be extended with an additional noise term, which yields the stochastic heat equation \cite{corwin2020some}. The noise term can be used to model chaotic microscopic effects and acts as a driving force for the system. Examples of these microscopic effects include molecular interactions or, as is the case in this work, neuronal noise.
%in this work, . 
%In this work, we use the noise term to model neuronal noise as a driving force of brain activity. 
%Noise is also ubiquitous in the brain and may play an important role in driving neurons 
%term can be added to this equation to model c. The noise term  which yields the stochastic heat equation. 

Both the heat equation and its stochastic extension are continuous in space. The heat equation can be naturally discretised by replacing the Laplacian operator with the negative graph Laplacian, yielding the graph heat equation \cite{xiao2009graph}.
%, which can not only be used to describe discrete Euclidean topologies, such as grids,
Crucially, the graph Laplacian can not only be used to describe discrete Euclidean topologies, such as grids, but also highly irregular graph topologies.
%Thanou et al. have used the graph heat equation to model multivariate signals residing on such irregular graphs using hidden heat sources \cite{thanou2017learning}.
%Brain signals typically reside on such irregular graphs due to the complex connectivity structure in the brain, the so-called connectome \cite{atasoy2016human}. 
Previous work has employed the graph heat equation to model multivariate signals using hidden heat sources learnt from the data \cite{thanou2017learning}.

In the current work, we combine the stochastic heat equation with the graph heat equation to model brain processing as a heat diffusion process driven by neuronal noise. 
On the one hand, this approach is motivated by the idea that multivariate neurophysiological signals reside on irregular graphs, which are a consequence of the complex connectivity structure in the brain \cite{atasoy2016human}.
On the other hand, it is underpinned by the widespread occurrence of neuronal noise in the brain and its potential role in driving neurons \cite{guo2018functional}.
%Multivariate signals typically reside on such irregular graphs due to the complex connectivity structure in the brain, the so-called connectome \cite{atasoy2016human}. 
Importantly, the additional stochastic term allows to use the signal values themselves as the heat sources, which has the principal advantage that they can be directly inferred from the data.
%While our work is based on Thanou et al. \cite{thanou2017learning}, the
%Using the introduced stochastic graph heat equation,

We use this model to develop a novel functional connectivity measure as a heat diffusion graph. 
To this end, we firstly solve the stochastic graph heat equation and adapt it to discrete signals with measurement noise. We then solve this equation for the graph Laplacian using suitable approximations.
Unlike common connectivity measures such as correlation, mutual information or Granger causality, which are bivariate \cite{bastos2016tutorial}, our diffusion-based connectivity measure is inherently multivariate. It further preserves the scale of the connectivity, which can be retrieved, for instance, as the spectral norm of the connectivity matrix. We refer to this scale as the \textit{graph thermal diffusivity}.

We empirically show that the graph thermal diffusivity is lower in electroencephalography (EEG) recordings of Alzheimer's disease (AD) patients compared to healthy controls, using two routine EEG datasets \cite{blackburn2018pilot,miltiadous2023dataset}.
Furthermore, we show that the graph thermal diffusivity correlates with Mini-Mental State Examination (MMSE) scores. 
We lastly link our results to previous findings in neuroimaging \cite{dai2014disrupted}.

\section{Theory}
\label{sec:theory}
\subsection{Stochastic Graph Heat Equation}
The heat equation is given by the following second-order differential equation:
\begin{align}
    \frac{\partial}{\partial t}x(s,t) = \Delta x(s,t)\label{eq:HE},
\end{align}
where $x(s, t)$ is a function, or field, in space and time and $\Delta$ is the Laplace operator. To model stochastic effects, a noise term can be added to equation \eqref{eq:HE}, yielding the stochastic heat equation:
\begin{align}
    \frac{\partial}{\partial t}x(s,t) = \Delta x(s,t) + \sigma\frac{\partial}{\partial t} W(s,t)\label{eq:SHE},
\end{align}
where $W(s,t)$ and $\sigma$ denote a Wiener process and its scale, respectively.
%Following work of Thanou et al. \cite{thanou2017learning}, w
We here extend equation \eqref{eq:SHE} to spatially discretised fields, i.d., time-continuous multivariate signals $\mathbf{x}(t)$ with an underlying spatial structure, which can be algebraically represented by an adjacency matrix $\mathbf{A}$.
To discretise equation \eqref{eq:SHE}, the continuous fields $x(s,t)$ and $W(s,t)$ are on the one hand replaced by the spatial signal $\mathbf{x}(t)$ and a vector of Wiener processes $\mathbf{W}(t)=\left(W_1(t), ..., W_N(t)\right)^\top$, respectively. On the other hand, the Laplace operator $\Delta$ is replaced by the negative graph Laplacian $\mathbf{L}=\mathbf{D}-\mathbf{A}$ \cite{thanou2017learning}, where $\mathbf{D}\coloneqq \mathrm{diag}(\mathbf{A}\cdot \mathbf{1})$ is the degree matrix, yielding overall:
\begin{align}
    \frac{\partial}{\partial t}\mathbf{x}(t) = -\mathbf{L} \mathbf{x}(t) + \sigma\frac{\partial}{\partial t} \mathbf{W}(t).\label{eq:partial}
\end{align}

\iffalse
extended to spatially discretised fields, i.d., multivariate signals $\mathbf{x}(t)$ continuous in time, whose spatial structures are irregular and are described by graphs. Specifically, the field $x(s, t)$ is replaced by the spatial signal $\mathbf{x}(t)$, while the Laplace operator $\Delta$ is given algebraically by the negative Laplacian matrix $\mathbf{L}$:
\begin{align}
    \frac{\partial}{\partial t}\mathbf{x}(t) = -\mathbf{L} \mathbf{x}(t).
\end{align}
\fi

\iffalse
The equation
\begin{align}
    \mathbf{x}(t) = e^{-t\mathbf{L}}\mathbf{x}_0
\end{align}
with the initial condition $\mathbf{x}(t=0) = \mathbf{x}_0$
solves xxx, which can be easily shown by insertion. 
%Here, $\mathbf{x}_0$ can be interpreted as a heat distribution which spreads
Thanou et al cite xxx interpreted the initial heat distribution $\mathbf{x}_0$ as an ensemble of hidden heat sources $\mathbf{h}^{(i)}_t$, which spread to the measured signal at time $t$ with different time scales $\tau_i$. The final measured signal $\mathbf{x}(t)$ is then given by:
\begin{align}
    \mathbf{x}(t) = \sum_{i=0}^S e^{-\tau_i \mathbf{L}}\mathbf{h}^{(i)}_t + \mathbf{e}_{ext},
\end{align}
where $\mathbf{e}_{ext} \sim \mathcal{N}_N(0, \sigma_{ext}\mathbb{1})$ is the measurement error.
\fi

\subsection{Solution of the Stochastic Graph Heat Equation}
\label{ssec:stochastic_graph_heat}
An exact solution for the stochastic graph heat equation \eqref{eq:partial} is given by:
\begin{align}
    \mathbf{x}(t) &= e^{-t\mathbf{L}}\left(\mathbf{x}_0 + \int_{t_0}^{t}e^{\tau\mathbf{L}}\sigma\frac{\partial}{\partial \tau}\mathbf{W}(\tau)d\tau\right), \label{eq:solution}
\end{align}
with the initial condition $\mathbf{x}(t_0) = \mathbf{x}_0$.
%\qquad t\ \mathrm{large\ enough},
The validity of this solution can be shown by inserting $\mathbf{x}(t)$ into equation \eqref{eq:partial}:
\begin{align}
    \frac{\partial}{\partial t} \mathbf{x}(t) = -&\mathbf{L} e^{-t\mathbf{L}}\left(\mathbf{x}_0 + \int_{t_0}^{t}e^{\tau\mathbf{L}}\sigma\frac{\partial}{\partial \tau}\mathbf{W}(\tau)d\tau\right)  \nonumber\\&+ e^{-t\mathbf{L}}e^{t\mathbf{L}}\sigma\frac{\partial}{\partial t}\mathbf{W}(t) \nonumber\\
    = -& \mathbf{L}\mathbf{x}(t) + \sigma \frac{\partial}{\partial t}\mathbf{W}(t).
\end{align}

\subsection{Model Sampling and Measurement Noise}
Equation \eqref{eq:solution} describes the continuous evolution of the signal in time at discrete spatial locations. To model an experimental recording, the signal can be sampled at equidistant time steps $\Delta t$ determined by the device sampling rate.
Generally, the time step $\Delta t$ is sufficiently small, such that the signal evolution can be approximated as follows:
%Sampling the solution xxx at small equidistant time steps $\Delta t$ yields approximately:
\begin{align}
    \mathbf{x}(t+\Delta t) =& e^{-(t+\Delta t)\mathbf{L}}\left(\mathbf{x}_0 + \int_{t_0}^{t+\Delta t}e^{\tau\mathbf{L}}\sigma\frac{\partial}{\partial \tau}\mathbf{W}(\tau)d\tau\right)  \nonumber\\=&e^{-\Delta t  \mathbf{L}}\left(\mathbf{x}(t) + \int_{t}^{t+\Delta t}e^{(\tau - t)\mathbf{L}}\sigma\frac{\partial}{\partial \tau}\mathbf{W}(\tau)d\tau\right) \nonumber\\
    \approx&e^{-\Delta t  \mathbf{L}}\left(\mathbf{x}(t) + \int_{t}^{t+\Delta t}\sigma\frac{\partial}{\partial \tau}\mathbf{W}(\tau)d\tau\right) \nonumber\\
    %=&e^{-\Delta t  \mathbf{L}}\left(\mathbf{x}(t) + \sigma\mathbf{W}(t+\Delta t)-\sigma\mathbf{W}(t)\right)\\
    =&e^{-\Delta t  \mathbf{L}}\left(\mathbf{x}(t) + \mathbf{e}\right),
\end{align}
where $\mathbf{e} \sim \mathcal{N}(\mathbf{0}, \sigma\Delta t)$ is a noise vector resulting from the definition of the Wiener process.

The signal $\mathbf{x}(t)$ corresponds to the system-internal signal. The external signal $\mathbf{x}'(t)$ includes measurement errors and is defined as:
\begin{align}
    \mathbf{x}'(t)\coloneqq & \mathbf{x}(t)+\mathbf{e}'\\
    \Rightarrow \mathbf{x}'(t+\Delta t) =& e^{-\Delta t  \mathbf{L}}\left(\mathbf{x}(t) + \mathbf{e}\right) + \mathbf{e}',\label{eq:simulation}
\end{align}
%\\ 
with $\mathbf{e}' \sim \mathcal{N}(\mathbf{0}, \sigma')$. The measurement error encompasses any external noise measured at the sensor which is not propagated by the system's graph connectivity.
Incidentally, equation \eqref{eq:simulation} can be used to simulate multivariate signal with an underlying spatial structure given by $\mathbf{L}$ \cite{goerttler2022effect}.

The sampled spatial signals $\mathbf{x}'(t_0 + k \Delta t)$, $k=0,...,N_c - 1$, can be concatenated to yield the full multivariate signal as a data matrix $\mathbf{X}'\in \mathbb{R}^{N_c \times N_t}$:
\begin{align}
    \mathbf{X}' = \begin{bmatrix}
        \mathbf{x}'(t_0),\mathbf{x}'(t_0 + \Delta t), ..., \mathbf{x}'(t_0 + (N_t - 1)\Delta t)
    \end{bmatrix}.
\end{align}
We further define data matrices $\mathbf{X}_0$ and $\mathbf{X}_1$ with index 0 or 1 by either removing the last column or the first column of $\mathbf{X}$, respectively. With these, the vector equation \eqref{eq:simulation} can be rewritten as a matrix equation:
\begin{align}
    \mathbf{X}_1' &= e^{-\Delta t  \mathbf{L}}\left(\mathbf{X}_0 + \mathbf{E}\right) + \mathbf{E}_1' \nonumber\\
    &=e^{-\Delta t  \mathbf{L}}\left(\mathbf{X}_0' - \mathbf{E}_0' + \mathbf{E}\right) + \mathbf{E}_1'\label{eq:matrix_eq},
\end{align}
where the noise matrices $\mathbf{E}\sim\mathcal{N}(\mathbf{0}_{N_c\times N_t},\sigma\Delta t)$ and $\mathbf{E}'\sim\mathcal{N}(\mathbf{0}_{N_c\times N_t},\sigma')$ are sampled from a multivariate normal distribution.

%To simulate the measurement error, a noise matrix $\mathbf{E}_{ext}\sim\mathcal{N}(\mathbf{0}_{N_c\times N_t},\sigma_{ext})$ is added to the internal system signal $\mathbf{X}$, yielding the recorded signal
%\begin{align}
%    \mathbf{X}_{rec}=\mathbf{X}+\mathbf{E}_{ext}
%\end{align}

%Note that this equation can be used to generate %... 

\iffalse
\subsection{Saturation effects}
The equation ... describes the heat model. However, the equation does not include saturation effects and measurement errors.
As $W(t)$ describes a Wiener process, the magnitude of the signal $\mathbf{x}(t)$ will increase over time. In order to model saturation effects, we bound the activity using the hyperbolic tangens as a sigmoid function with scaling parameter $\beta$:
\begin{align}
    \mathbf{x}(t+\Delta t) = \beta\tanh\left(\frac{1}{\beta}e^{-\Delta t\mathbf{L}}(\mathbf{x}(t) + \mathbf{e})\right).
\end{align}
\fi

%\subsection{}

\section{Methods}
\label{sec:methods}

\subsection{Heat Diffusion Graph Retrieval}
This section shows how an algebraic heat graph can be retrieved from a multivariate signal governed by heat-like graph dynamics described in section \ref{ssec:stochastic_graph_heat}.
%Firstly, two data matrices $\mathbf{X}_0$ and $\mathbf{X}_1$ are extracted from $\mathbf{X}$ by either removing the last column or the first column of $\mathbf{X}$, respectively.
%Crucially, the vector equation for spatial signals \eqref{eq:simulation} can be easily extended for the two matrices as follows:
%\begin{align}
%    \mathbf{X}_1 = e^{-\Delta t \mathbf{L}}(\mathbf{X}_0 + \mathbf{E}_{0}),
%\end{align}
%where $\mathbf{E}_{0} \sim\mathcal{N}(\mathbf{0}_{N_c\times N_t-1}, \sigma\Delta t)$
%\begin{align}
%    \mathbf{E}_{0} = [\mathbf{e}_{t_0}, ..., \mathbf{e}_{t_{N_t - 2}}]\sim\mathcal{N}(\mathbf{0}_{N_c\times N_t-1}, \sigma\Delta t)
%\end{align} 
%is the internal noise data matrix.
%If measurement noise is present, $\mathbf{X}$ is replaced with $\mathbf{X}_{rec}-\mathbf{E}_{ext}$, yielding
The principal goal is to retrieve the graph Laplacian $\mathbf{L}$ from equation \eqref{eq:matrix_eq}.
%, given by:
%\begin{align}
%    \mathbf{X}_1' &=e^{-\Delta t  \mathbf{L}}\left(\mathbf{X}_0' - \mathbf{E}_0' + \mathbf{E}\right) + \mathbf{E}_1'.
%\end{align}
%\begin{align}
%    \mathbf{X}_{1}'-\mathbf{E}_{1}'= e^{-\Delta t \mathbf{L}}(\mathbf{X}_{0}' - \mathbf{E}_{0}' + \mathbf{E}_{0}).\label{eq:matrix_equation}
%\end{align}
%In the following, the goal is to retrieve the Laplacian matrix $\mathbf{L}$ from this equation.
Note that generally only the recorded signal $\mathbf{X}'$ is given. However, assumptions about $\sigma\Delta t$ and $\sigma'$ as well as noise term averages allow to infer the graph structure and the scale of the heat processing.

To simplify our calculations, we begin by defining the following two matrices:
\begin{align}
    \mathbf{M}_1 &= \mathbf{X}_{1}'-\mathbf{E}_{1}'\\
    \mathbf{M}_0 &= \mathbf{X}_{0}' - \mathbf{E}_{0}' + \mathbf{E}_{0}.
\end{align}
We then insert these two definitions into equation \eqref{eq:matrix_eq} and solve the equation for $\mathbf{L}$, yielding:
\begin{align}
    &&\mathbf{M}_1 &= e^{-\Delta t \mathbf{L}}\mathbf{M}_0\\
    \Leftrightarrow && \mathbf{M}_1 \mathbf{M}_0^\top (\mathbf{M}_0 \mathbf{M}_0^\top)^{-1} &= e^{-\Delta t \mathbf{L}}\\
    \Leftrightarrow && \mathbf{L} = -\frac{1}{\Delta t}\log &\left((\mathbf{M}_1 \mathbf{M}_0^\top) (\mathbf{M}_0 \mathbf{M}_0^\top)^{-1} \right).\label{eq:L_formula}
\end{align}
We further make use of the fact that a noise matrix multiplied with any matrix other than itself averages to zero, and that for small enough time steps $\mathbf{X}_{1}'$ contains the noise matrix $\mathbf{E}_{0}$. This yields the following approximations:
\begin{align}
    \mathbf{M}_1 \mathbf{M}_0^\top &\approx \mathbf{X}_{1}' \mathbf{X}_{0}'^\top + \mathbf{X}_{1}'  \mathbf{E}_{0}^\top \nonumber\\
    &\approx \mathbf{X}_{1}' \mathbf{X}_{0}'^\top + \mathbf{E}_{0} \mathbf{E}_{0}^\top\label{eq:M1M0}\\
    \mathbf{M}_0 \mathbf{M}_0^\top &\approx \mathbf{X}_{0}'\mathbf{X}_{0}'^\top + \mathbf{E}_{0}'\mathbf{E}_{0}'^{\top} + \mathbf{E}_{0}\mathbf{E}_{0}^{\top}\label{eq:M0M0}. 
\end{align}
To estimate the remaining noise terms, we observe that for small $\Delta t$ equation \eqref{eq:matrix_eq} reduces to
%\begin{align}
    $\mathbf{X}_{1}' - \mathbf{E}_{1}' \approx \mathbf{X}_{0}' - \mathbf{E}_{0}' + \mathbf{E}_{0}$,
%\end{align}
such that:
\begin{align}
    \left(\mathbf{X}_{1}' - \mathbf{X}_{0}'\right)\left(\mathbf{X}_{1}' - \mathbf{X}_{0}'\right)^{\top}  \approx \mathbf{E}_{0}' \mathbf{E}_{0}'^{\top} + \mathbf{E}_{0} \mathbf{E}_{0}^{\top} + \mathbf{E}_{1}' \mathbf{E}_{1}'^{\top}.
\end{align}
Without prior knowledge about the standard deviations $\sigma\Delta t$ and $\sigma'$, we may assume $\sigma\Delta t \approx \sigma'$, allowing us to estimate: 
\begin{align}
    \mathbf{E}_{0}'\mathbf{E}_{0}'^{\top} &\approx \mathbf{E}_{0} \mathbf{E}_{0}^{\top} \approx \mathbf{E}_{1}' \mathbf{E}_{1}'^{\top}  \nonumber\\ &\approx \frac{1}{3}\left(\mathbf{X}_{1}' - \mathbf{X}_{0}'\right)\left(\mathbf{X}_{1}' - \mathbf{X}_{0}'\right)^{\top} \eqqcolon \tilde{\mathbf{E}}\tilde{\mathbf{E}}^{\top}.\label{eq:approximation}
\end{align}
Inserting these approximations into equations \eqref{eq:M1M0} and \eqref{eq:M0M0} yields:
\begin{align}
    \mathbf{M}_1 \mathbf{M}_0^\top 
    &\approx \mathbf{X}_{1}' \mathbf{X}_{0}'^\top + \tilde{\mathbf{E}} \tilde{\mathbf{E}}^\top\\
    \mathbf{M}_0 \mathbf{M}_0^\top &\approx \mathbf{X}_{0}'\mathbf{X}_{0}'^\top + 2\tilde{\mathbf{E}}\tilde{\mathbf{E}}^{\top}.%\\
    %\mathbf{M}_0 \mathbf{M}_0^\top &\approx \mathbf{X}_{0}'\mathbf{X}_{0}'^\top + \frac{2}{3}\left(\mathbf{X}_{1}' - \mathbf{X}_{0}'\right)\left(\mathbf{X}_{1}' - \mathbf{X}_{0}'\right)^{\top},
\end{align}
These approximations can finally be inserted into equation \eqref{eq:L_formula}, such that the graph Laplacian $\mathbf{L}$ is given solely in terms of the recorded data matrix $\mathbf{X}'$ and the sampling resolution $\Delta t$:
\begin{align}
    \mathbf{L} \approx -\frac{1}{\Delta t}\log \biggl(&\Bigl(\mathbf{X}_{1}' \mathbf{X}_{0}'^\top + \frac{1}{3}\left(\mathbf{X}_{1}' - \mathbf{X}_{0}'\right)\left(\mathbf{X}_{1}' - \mathbf{X}_{0}'\right)^{\top}\Bigr) \nonumber\\ &\Bigl(\mathbf{X}_{0}'\mathbf{X}_{0}'^\top + \frac{2}{3}(\mathbf{X}_{1}' - \mathbf{X}_{0}')(\mathbf{X}_{1}' - \mathbf{X}_{0}')^{\top}\Bigr)^{-1} \biggr)\label{eq:heat_L}.
\end{align}

\subsection{Graph Laplacian Constraints}
In general, the graph Laplacian computed in equation \eqref{eq:heat_L} does not yet satisfy all the properties of a Laplacian corresponding to an undirected, non-negative graph $\mathbf{A}$, which are: 
\begin{align}
    (i)\phantom{ii}&& \mathrm{diag}(\mathbf{L}) &= \mathbf{A}\cdot \mathbf{1},&&\\
    (ii)\phantom{i}&&\mathbf{A} &= \mathbf{A}^\top,&&\\
    (iii)&&[\mathbf{A}]_{ij} &\geq 0.&&
\end{align}
By imposing these constraints on $\mathbf{L}$, we can further improve the retrieval of the graph Laplacian $\mathbf{L}$ in equation \eqref{eq:heat_L} . To enforce properties (ii) and (iii), we simply symmetrise the adjacency matrix and set all negative elements to zero:
\begin{align}
    \mathbf{A}_{s} &= \frac{\mathbf{A}+\mathbf{A}^\top}{2},\\
    [\mathbf{A}_{s+}]_{ij} &= \max([\mathbf{A}_{s}]_{ij}, 0).
\end{align}
To enforce constraint (i), we firstly compute the degree matrix $\mathbf{D}_{s+} = \mathrm{diag}(\mathbf{A}_{s+} \cdot \mathbf{1})$. To include the information given in the measured degree matrix $\mathbf{D}=\mathbf{L} - \mathbf{A}$, we mix $\mathbf{D}_{s+}$ with $\mathbf{D}$ to yield the diagonal degree matrix $\tilde{\mathbf{D}}$:
\begin{align}
    [\tilde{\mathbf{D}}]_{ii} = \sqrt{\frac{\max([\mathbf{D}]_{ii},0) + [\mathbf{D}_{s+}]_{ii}}{2 [\mathbf{D}_{s+}]_{ii}}},
\end{align}
%where $\tilde{\mathbf{D}}$ is the resulting diagonal degree matrix.
Note that we limited $[\mathbf{D}]_{ii}$ to non-negative values to ensure that the argument of the root is positive. 
Finally, the adjacency matrix satisfying all three properties is given by:
\begin{align}
    \tilde{\mathbf{A}} = \tilde{\mathbf{D}}^{1/2}\mathbf{A}_{s+}\tilde{\mathbf{D}}^{1/2},
\end{align}
from which the final graph Laplacian can be inferred as $\tilde{\mathbf{L}}=\mathrm{diag}(\tilde{\mathbf{A}}\cdot \mathbf{1}) - \tilde{\mathbf{A}}$.

\begin{figure*}[thpb]
  \centering
  \includegraphics[width=0.99\textwidth]{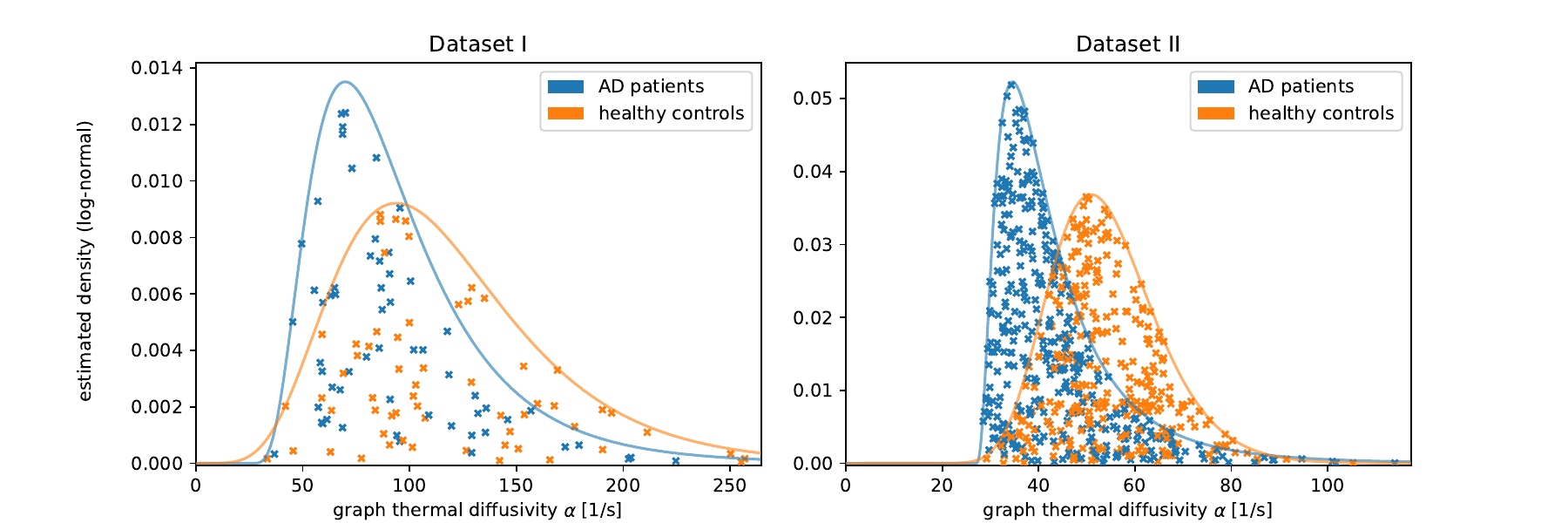}
  %\begin{tikzpicture}
  %\sffamily
  %  \node[anchor=south west,inner sep=0] at (0,0) {\includegraphics[width=0.99\textwidth]{figures/graph_thermal_diffusivity_Alzheimers_alpha_sd_ld.pdf}};
   %     \node (caption) at (1.8,5.5){A};
   %     \node (caption) at (9.2,5.5){B}; 
  %\end{tikzpicture}
  \caption{Graph thermal diffusivity distributions for dataset I (left) and dataset II (right) for AD patients (blue) and healthy controls (orange) using a propagation time of $\Delta t=27.5\,\mathrm{ms}$. 
  %, corresponding, e.g., to a propagation distance of $\Delta x\approx30\,\mathrm{mm}$ for alpha rhythms with a measured speed of $v\approx910\,\mathrm{mm}$ \cite{halgren2019generation}
  The distributions were modelled as log-normal distributions.
  %due to the boundedness of the diffusivity. 
  Crucially, the distributions for the AD patients are shifted towards lower diffusivity values for both datasets, indicative of a lower brain activity spread.
  %, which indicates that brain activity does not spread, or diffuse, as effectively for AD patients as compared to healthy controls. 
  %This in turn suggests that the structural connectivity for AD patients is impaired on this time scale, which is in line with previous findings \cite{dai2014disrupted}. 
  Note also that the locations of the distributions are higher for dataset I.}
  %, likely due to differences in the montage and the length}
  %, which may be due to differences in the EEG setup and montage as well as differences in the signal preprocessing}
  \label{fig:diffusivity}
\end{figure*}

\subsection{Graph Thermal Diffusivity}
%Our proposed method 
%Unlike other connectivity measures, our proposed method 
%xxx
%
%Our proposed method does not only estimate the graph structure in the multivariate signal, but also a scale of the connectivity in terms of dynamics.
Our proposed method estimates a graph structure in the multivariate signal scaled in terms of its dynamic evolution.
%This is different from other functional connectivity measures such as
To explicitly retrieve this scale, we normalise the final Laplacian matrix by its spectral norm $\lVert\tilde{\mathbf{L}}\rVert_2=\lambda_{max}(\tilde{\mathbf{L}})$, such that $\alpha\coloneqq \lambda_{max}(\tilde{\mathbf{L}})$ is the scale of heat-like processing in the data matrix, given in units of $1/\mathrm{s}$. In analogy with the classical heat model, we term $\alpha$ the \textit{graph thermal diffusivity}.

\section{Experiments}
\label{sec:results}
%\subsection{Alzheimer's disease}
\subsection{Dataset I}
The first routine EEG dataset used in this study was recorded by Blackburn et al. on 20 AD patients and 20 healthy controls \cite{blackburn2018pilot}. 
Participants were instructed to close their eyes during the recording.
A combination of the 10-20 and the 10-10 electrode placement system was used, from which 23 bipolar channels were created. 
The sampling rate was $2048\,\mathrm{Hz}$. 
%We further downsampled the data to a rate of $204.8\,\mathrm{Hz}$. 
For each patient, two or three segments with a length of 12 seconds were selected from the recording, resulting in overall 119 samples.
In addition, MMSE scores were obtained from 13 out of the 20 AD patients. 
% with shape $N_c\times N_t= 23\times 24,576$. 
%More details about the dataset can be found in Blackburn et al. \cite{blackburn2018pilot}.

%The second, publicly available\footnote{\href{https://openneuro.org/datasets/ds004504/versions/1.0.6} {https://openneuro.org/datasets/ds004504/versions/1.0.6}} dataset comprises 36 Alzheimer's disease patients and 29 healthy controls \cite{miltiadous2023dataset}.
%It was recorded using an EEG system with 19 scalp electrodes and two reference electrodes at a sampling rate of 500$\,$Hz. As in dataset I, patients were instructed to close their eyes during the measurement. Measurements lasted for roughly 13-14 minutes on average. The EEG samples were filtered and re-referenced using the reference electrodes, and artefact detection methods were employed. More details about the dataset and the preprocessing are described in Miltiadous et al. \cite{miltiadous2023dataset}.
%We partitioned the preprocessed EEG samples into 60 seconds-long samples, excluding samples that contained artefacts. This resulted in overall 656 EEG samples with shape $N_c \times N_t=19\times 30,000$.

\subsection{Dataset II}
The second dataset is publicly available\footnote{\href{https://openneuro.org/datasets/ds004504/versions/1.0.6} {https://openneuro.org/datasets/ds004504/versions/1.0.6}} and was recorded on 36 AD patients and 29 healthy controls \cite{miltiadous2023dataset}. 
It was acquired using an EEG system with 19 scalp electrodes placed according to the 10-20 system and two reference electrodes, sampled at a rate of 500$\,$Hz.
%The EEG data used in this study was recorded from 20 Alzheimer's patients and 20 healthy controls at a sampling rate of $2048\,\mathrm{Hz}$ using 23 channels, which were transformed to create 23 bipolar channels. 
%We further downsampled the data to a rate of $204.8\,\mathrm{Hz}$. 
Similar to dataset I, participants were directed to close their eyes throughout the recordings, which lasted for approximately 13-14 minutes on average. 
MMSE scores were assessed for each AD patient.
%The obtained EEG samples underwent filtering and re-referencing procedures utilizing the reference electrodes, and artifact detection methods were applied. 
Additional information regarding the dataset and preprocessing steps can be found in Miltiadous et al. \cite{miltiadous2023dataset}.
We partitioned the preprocessed EEG recordings into segments lasting 60 seconds each. We further excluded segments containing artefacts, resulting in a total of 656 samples. 

\subsection{Experiment Design}
In our experiment, we firstly applied a Butterworth bandpass filter between 0.5\,$\mathrm{Hz}$ and 45\,$\mathrm{Hz}$ to filter out noise.
We then computed the graph thermal diffusivity $\alpha$ for each sample in the two datasets using a propagation time of $\Delta t= 27.5\,\mathrm{ms}$, which we set by downsampling datasets I and II by factors of 56 and 13, respectively. 
This propagation time is informed by typical brain signal propagation speeds;
%, which are on the order of $v=1\,\mathrm{m}/\mathrm{s}$. 
for instance, Halgren et al. measured the alpha rhythm propagation in human cortex as $v=0.91\,\mathrm{m}/\mathrm{s}$ \cite{halgren2019generation}, which would correspond to a propagation distance of $\Delta x\approx 25\,\mathrm{mm}$ for a time step of $\Delta t= 27.5\,\mathrm{ms}$. 

%Brain signal propagation is typically much slower than the EEG device's time resolution, which is why we downsampled both datasets to a time resolution of $\Delta t\approx 30\,\mathrm{ms}$.
%Such a time scale 
%In order to capture brain signal diffusion, we defined propagation times... \cite{halgren2019generation} xxx

\subsection{Results}
Figure \ref{fig:diffusivity} shows the retrieved diffusivity $\alpha$ values for AD patients in blue and healthy controls in orange. We fitted log-normal distributions, suitable due to their boundedness at zero and good fit, to the results using the Python distfit package \cite{Taskesen_distfit_is_a_2020}. The values are systematically higher for dataset I, possibly reflecting differences in the EEG montage and segment length. Most notably, the graph thermal diffusivity distributions are consistently shifted towards lower values for AD patients across both datasets. 
The Pearson correlation coefficient between the MMSE score and the graph thermal connectivity is marginal at $\rho=0.17$ ($p<0.1$) for dataset I, but highly significant at $\rho=0.35$ ($p<0.001$) for dataset II.

%For dataset II, we are able to 
%.locally restricted brain processing, possibly as a consequence of structural impairment.
%In particular, a lack of lower diffusivity values points at an inability to overcome heat-like signal processing, which we assume to be a low-level form of brain processing. On the other hand, a lack of higher diffusivity values indicates more locally restricted brain processing, possibly as a consequence of structural impairment.

\section{Discussion}
\label{sec:discussion}
In this study, we have introduced a novel connectivity measure based on heat diffusion-like signal propagation. The connectivity measure is inherently multivariate and yields a scaling factor, the so-called graph thermal diffusivity.
In our EEG-based experiments, we have observed a shift towards lower diffusivity values in AD patients, which indicates that the brain processing may be more local. We therefore interpret the increase of low diffusivity values in AD patients as a consequence of structural impairment in those patients \cite{dai2014disrupted}.
In addition, we have found a highly significant correlation between the MMSE score and the graph thermal diffusivity for the larger dataset II, giving further evidence for the link between AD and the graph thermal connectivity.
We believe that the diffusivity may contribute to EEG-based AD diagnosis as an additional classification feature.

Our method has several limitations. Firstly, it involves several approximations, which may not be valid in every setting.
Secondly, our method requires the computation of a matrix inverse, which may lead to unreliable results if the matrix is ill-conditioned. Lastly, the simplicity of the underlying model may not adequately capture the complexity of EEG data and is limited to a specific time scale. Despite these challenges, our experimental AD characterisation results showcase the value of our heat diffusion connectivity measure and its scaling factor, the graph thermal diffusivity. Due to the simplicity of the model, our method may be applicable to a broader range of problems involving multivariate signals.

\addtolength{\textheight}{-0cm}   % This command serves to balance the column lengths

%%%%%%%%%%%%%%%%%%%%%%%%%%%%%%%%%%%%%%%%%%%%%%%%%%%%%%%%%%%%%%%%%%%%%%%%%%%%%%%%

%%%%%%%%%%%%%%%%%%%%%%%%%%%%%%%%%%%%%%%%%%%%%%%%%%%%%%%%%%%%%%%%%%%%%%%%%%%%%%%%

%%%%%%%%%%%%%%%%%%%%%%%%%%%%%%%%%%%%%%%%%%%%%%%%%%%%%%%%%%%%%%%%%%%%%%%%%%%%%%%%
%\section*{APPENDIX}

%\section*{Acknowledgment}
%assf
%The preferred spelling of the word ``acknowledgment" in America is without an ``e" after the ``g". Avoid the stilted expression, ``One of us (R. B. G.) thanks . . ."  Instead, try ``R. B. G. thanks". Put sponsor acknowledgments in the unnumbered footnote on the first page.

%%%%%%%%%%%%%%%%%%%%%%%%%%%%%%%%%%%%%%%%%%%%%%%%%%%%%%%%%%%%%%%%%%%%%%%%%%%%%%%%

%References are important to the reader; therefore, each citation must be complete and correct. If at all possible, references should be commonly available publications.

%\begin{thebibliography}{99}
%\end{thebibliography}

\bibliographystyle{IEEEtran}
\bibliography{refs}

\end{document}